\documentstyle[12pt]{article}
\begin{document}
\tolerance=5000
\def\be{\begin{equation}}
\def\ee{\end{equation}}
\def\bea{\begin{eqnarray}}
\def\eea{\end{eqnarray}}
\def\nn{\nonumber \\}
\def\cF{{\cal F}}
\def\det{{\rm det\,}}
\def\Tr{{\rm Tr\,}}
\def\e{{\rm e}}
\def\etal{{\it et al.}}
\def\erp2{{\rm e}^{2\rho}}
\def\erm2{{\rm e}^{-2\rho}}
\def\er4{{\rm e}^{4\rho}}
\def\etal{{\it et al.}}
\def\gsim{\ ^>\llap{$_\sim$}\ }

\  \hfill
\begin{minipage}{3.5cm}
NDA-FP-49 \\
June 1998 \\
\end{minipage}

\begin{center}

{\large\bf Quantum generation \\
of Schwarzschild-de Sitter (Nariai) black holes \\
in effective dilaton-Maxwell gravity}

\vfill

{\sc Andrei A. BYTSENKO}\footnote{e-mail:
abyts@fisica.uel.br},
{\sc Shin'ichi NOJIRI$^{\clubsuit}$}
\footnote{e-mail : nojiri@cc.nda.ac.jp} \\
and {\sc Sergei D. ODINTSOV$^{\spadesuit}$}
\footnote{e-mail : odintsov@tspi.tomsk.su}

{\sl Sankt-Petersburg Technical
University, RUSSIA \\
and \\
Departamento de Fisica, Universidade Estadual de Londrina, \\
Caixa Postal 6001, Londrina-Parana, BRAZIL} \\

{\sl $\clubsuit$
Department of Mathematics and Physics National Defence Academy, \\
Hashirimizu Yokosuka 239, JAPAN} \\

{\sl $\spadesuit$
Tomsk Pedagogical University, 634041 Tomsk, RUSSIA} \\

\vfill

{\bf ABSTRACT}

\end{center}

Dilaton coupled electromagnetic field is essential element
of low-energy string effective action or it may be considered as
result of spherical compactification of Maxwell theory in higher dimensions.
The large $N$ and large curvature effective action for $N$ dilaton coupled
vectors is calculated. Adding such quantum correction to classical
dilaton gravity action we show that effective dilaton-Maxwell gravity
under consideration may generate Schwarzschild-de Sitter black
holes (SdS BHs) with constant dilaton
 as solutions of the theory. That suggests a mechanism
(alternative to BHs production) for quantum generation of SdS BHs in
early universe  (actually, for quantum creation of inflationary Universe)
 due to back-reaction of dilaton coupled matter. The possibility of
proliferation of anti-de Sitter space is briefly discussed.

\


\newpage

In the recent paper by Bousso\cite{B} it has been presented the evidence
that Schwarzschild-de Sitter (SdS) black holes (BHs) in their extremal
limit which is known as Nariai BH may be responsible for proliferation of
de Sitter universe (perdurence of de Sitter space has been discussed long
ago by Ginsparg and Perry\cite{GP}). In other words, in the process
of quantum evolution of SdS (Nariai) BHs, de Sitter universe may
disintegrate to the number of de Sitter universes (proliferation of de
Sitter space). If such process really exists, it could lead to
various interesting cosmological consequences mentioned in ref.\cite{B}.

It is known that SdS BH may be obtained as the solution of classical
Einstein gravity with cosmological constant term. It does not appear as
the result of star collapse, rather it may be created at the
initial (inflationary) stage of the universe. (Roughly speaking, SdS BH
may be described as de Sitter space with BH immersed inside of it).
As quantum gravitational effects are relevant in this stage, the
interesting question is : can SdS BH be generated by quantum effects in
the early universe? In the present letter we try to give the answer to
this question, working with specific model of $N$ dilaton coupled
vector fields. We calculate the effective Lagrangian for that model in
large $N$ approximation (what justifies the neglecting of
quantum gravitational and dilaton contributions) and study corresponding
quantum corrected equations of motion. We show that indeed SdS
(Nariai) solution of such equations exists. (On the classical level it
was not possible). That gives the alternative way to generate SdS BH
by quantum effects of dilaton coupled matter in early universe. In
the similar way S(anti)dS BHs may be generated.

We start from the conformally invariant
 action of dilaton coupled
vector field in curved background
 \be \label{I} S_{EM}=-{1 \over 4}\int
d^4x \sqrt{-g}f(\phi) F^a_{\mu\nu}F^{a\mu\nu} \ee where $\phi$ is
dilaton, $f(\phi)$ is an arbitrary dilatonic function, $F^a_{\mu\nu}$
is the strength for  $N$ vector fields. The action (\ref{I}) describes
$N$ dilaton coupled quantum electromagnetic fields \cite{NO} (see also
\cite{IO}). In its own turn, dilaton coupled electromagnetic field is
the essential element of the low energy string effective action:  \be
\label{II}
S=\int d^4x\sqrt{-g}\left[ R + 4 (\nabla \phi)^2
+ F_{\mu\nu}^2 \right]\e^{-2\phi}\ .
\ee

Dilaton coupled electromagnetic field may be obtained also
from spherical compactification of higher dimensional Maxwell theory.
Our purpose will be the study of  vacuum renormalization of the theory
with the action (\ref{I}). According to general arguments
(see \cite{BOS}, for a review), quantum field theory in an
external background can be made renormalizable (if it was
renormalizable without such background) and if the proper
external fields action and non-minimal coupling type action
are added to the original action. Due to gauge invariance,
the non-minimal type terms do not appear in our problem.
The theory is free (quadratic) one. However, there is non-trivial
vacuum structure.

The external fields action may be written as following
\cite{NO}
\be
\label{III}
S_{ext}=\int d^4x \sqrt{-g}\sum_{i=1}^{13}a_iJ_i
\ee
where $a_i$ are vacuum coupling constants, $J_1=F$ is
the square of Weyl tensor, $J_2=G$ is Gauss-Bonnet
invariant, $J_3=\Box R$, $J_4=\Box\left(f^{-1}\Box f\right)$,
$J_5=\Box\left( f^{-2}(\nabla_\alpha f)(\nabla^\alpha f)\right)$,
$J_6=f^{-2}R_{\mu\nu}(\nabla^\mu f)(\nabla^\nu f)$,
$J_7=f^{-1}R^{\mu\nu}\nabla_\mu\nabla_\nu f$,
$J_8=R f^{-1}\Box f$,
$J_9=f^{-4}(\nabla_\mu f)(\nabla^\mu f)
(\nabla_\alpha f)(\nabla^\alpha f)$,
$J_{10}=f^{-3}(\nabla^\beta f)(\nabla^\nu f)
(\nabla_\beta\nabla_\nu f)$,
$J_{11}=f^{-3}(\nabla_\mu f)(\nabla^\mu f)(\Box f)$,
$J_{12}=f^{-2}(\nabla_\alpha\nabla_\beta f)
(\nabla^\alpha\nabla^\beta f)$, $J_{13}=(\Box f)(\Box f)$.
Note that $R^2$ term does not appear independently as the
counterterm due to conformal invariance. It appears only
as part of $F$ and $G$ in above model.

It is not difficult to analyze the theory with the action
\be
\label{IV}
S=S_{EM}+S_{ext}
\ee
and to show that such a theory is trivially renormalizable one.
This theory represents just usual free quantum field theory
in curved spacetime with external dilaton. We take the number of
vectors $N$ to be large enough in order to consider above model
as some effective theory for quantum dilaton-Maxwell gravity.
Due to use of large $N$ approximation one can neglect quantum
effects of dilaton and graviton.

 Let us discuss now the one-loop renormalization (whole
renormalization is defined by only one-loop renormalization). The
calculation of the corresponding one-loop effective action may be done
using standard background field method (see \cite{BOS}, for a review).
Using explicit form of the one-loop effective action, one can easily
find $\beta$-functions for all coupling constants (vacuum
$\beta$-functions have been actually found in \cite{NO,IO}):
\bea
\label{V}
&& \beta_{a_1}={N \over 10(4\pi)^2}\ , \ \
\beta_{a_2}=-{31N \over 180(4\pi)^2}\ , \ \
\beta_{a_3}=-{N \over 10(4\pi)^2}\ , \nn
&& \beta_{a_4}=-{N \over 30(4\pi)^2}\ , \ \
\beta_{a_5}={5N \over 12(4\pi)^2}\ , \ \
\beta_{a_6}=-{N \over 3(4\pi)^2}\ , \nn
&& \beta_{a_7}=-{N \over 3(4\pi)^2}\ , \ \
\beta_{a_8}={N \over 6(4\pi)^2}\ , \ \
\beta_{a_9}={9N \over 16(4\pi)^2}\ , \nn
&& \beta_{a_{10}}={N \over 6(4\pi)^2}\ , \ \
\beta_{a_{11}}=-{11N \over 12(4\pi)^2}\ , \ \
\beta_{a_{12}}=-{N \over 6(4\pi)^2}\ , \nn
&& \beta_{a_{13}}={5N \over 12(4\pi)^2}\ .
\eea
Here $N$ is the number of components of gauge field:
$a=1, \cdots, N$. Above RG functions are defined as
standard vacuum $\beta$-functions for quantum field theory
in curved spacetime \cite{BOS}.

With above $\beta$-functions, it is not difficult to construct
the effective coupling constants:
\be
\label{VI}
a_i(t)=\tilde a_i + \beta_{a_i}t\ ,\ \
\ee
where $\tilde a_i = a_i(t=0)$.

Due to renormalizability of the theory, the
effective Lagrangian satisfies the following vacuum RG equation
\be
\label{VII}
\left(\mu{\partial \over \partial \mu}
+\beta_{a_i}{\partial \over \partial a_i}
\right)
L_{eff}(\mu, a_i, g_{\mu\nu}, \phi)
=0\ .
\ee
Let us note that the form of RG equation is the same as
in the absence of dilaton background \cite{BOS}.

We work in the background field method \cite{BOS} where
 one can
represent the effective Lagrangian in the following form
\be
\label{VIII}
L_{eff}=L_{ext}(\mu, a_i, g_{\mu\nu}, \phi)
\ee
Here  $L_{ext}$ is the
effective action for external fields.
Applying RG Eq.(\ref{VII}) to Eq.(\ref{VIII}), we
may solve it as
\be
\label{XI}
L_{ext}(\mu, a_i, g_{\mu\nu}, \phi)
=L_{ext}(\mu\e^t, a_i(t),  g_{\mu\nu}, \phi)\ .
\ee
Here ${d a_i(t) \over dt}=\beta_{a_i}(t)$.
Hence, improved effective Lagrangian for dilaton coupled
EM theory in curved space is given by above equation.
(For scalar self-interacting theory in flat space, see refs.\cite{CW}).

Working in the leading approximation, one should use some
boundary condition to define the effective Lagrangian at
$t=0$.
As such boundary condition one can use the tree-level action,
i.e., sum of Eqs.(\ref{I}) and (\ref{III}).
Then we obtain
\be
\label{XII}
L_{eff}= \sum_{i=1}^{13}a_i(t)J_i\ .
\ee

One has to think now about the meaning of RG parameter $t$,
which normally corresponds to the logarithm of the effective
mass squared.

Let us consider the case of strong gravitational field. Then,
the effective mass naturally defines RG parameter $t$ to be:
\be
\label{XIV}
t={1 \over 2}\ln {|f(\phi) R| \over \mu^2}\ .
\ee
This expression should be used in Eq.(\ref{XII}) in order to
give $L_{eff}$ at strong values of $f(\phi)R$. Note that if
one calculates the effective action at strong curvature directly
(without use of RG) the result would be just the same. The only
difference may enter through initial values of effective coupling
constants which should be defined by regularization. However, at
strong curvature this difference is not relevant.

Let us discuss now some BH-type solutions which follow from
$L_{eff}$ (\ref{XII}).
For simplicity, we consider the solution where the electromagnetic fields
vanish. Then we can start with the following
action (where first part may be considered as classical
action for dilaton-Maxwell gravity).
\be
\label{XVII}
S={1 \over 16\pi G}\int d^4x\sqrt{-g}\left[ R
 + 4 (\nabla \phi)^2  + 4\lambda^2 \right]\e^{-2\phi}
+\int d^4x \sqrt{-g}\sum_{i=1}^{13}a_i(t)J_i
\ee
and we also assume
\be
\label{f}
f(\phi)=\e^{-2\phi}
\ee
Note that we neglect the kinetic term
for the electromagnetic fields since this term does not give any contribution
to the equations of motion for solutions with vanishing
electromagnetic fields.
Since we are now interested in the static and spherical solution, we
reduce the action (\ref{XVII}) into two dimensional one by
assuming the metric to be
\be
\label{XVIII}
ds^2=\sum_{\mu,\nu=0,1}g_{\mu\nu}dx^\mu dx^\nu + \e^{-2\sigma}d\Omega^2\ .
\ee
Here $d\Omega^2$ is the line element on two dimensional sphere $S^2$.
We also consider the solution where $\phi$ and $\rho$ are constants.
This solution does not exist in classical theory when there is
no cosmological term. One specific type of such solution is known as
the Nariai solution and its quantum (in)stability was studied by
Bousso-Hawking \cite{BH2} and authors \cite{NO2}. Recently Bousso has
shown the possibility that de Sitter space proliferates by using this
solution. Therefore if such a solution exists when the scalar curvature
is large at the early universe, such a solution would
affect the cosmology.

Then 2d reduced effective action becomes
\bea
\label{XIVb}
{S \over 4\pi}&=&{1 \over 16\pi G}\int dx^2\sqrt{-g}\e^{-2\sigma-2\phi}
\left({\cal R}+2\e^{2\sigma} + 4\lambda^2 \right) \nn
&& + \int d^2x\sqrt{-g}\e^{-2\sigma}
\left\{{a_1(t) \over 3}\left({\cal R}+2\e^{2\sigma}\right)^2
+ 4a_2 \e^{2\sigma}{\cal R}\right\} + \cdots
\eea
Here ${\cal R}$ is the scalar curvature in two dimensions and
``$\cdots$" expresses the terms containing the second power of
the derivatives of ${\cal R}$, $\phi$ or $\rho$, which do not contribute
to the equations of motion if $\sigma$ and $\phi$ are constant.
Then neglecting the derivatives over ${\cal R}$, $\phi$ or $\rho$,
the equations of motion given by the variations over $\phi$, $\sigma$
and $\rho$ have the following form in the
conformal gauge $g_{\pm\mp}=-{1 \over 2}\e^{2\rho}\ ,\ \ g_{\pm\pm}=0$:
\bea
\label{phieq}
0&=&-{1 \over 16\pi G}\e^{-2\phi}\left({\cal R}+2\e^{2\sigma}
+4 \lambda^2 \right)  \nn
&& - {1 \over 2}\left\{{\beta_{a_1} \over 3}\left({\cal R}
+2\e^{2\sigma}\right)^2 + 4\beta_{a_2}\e^{2\sigma}{\cal R}\right\} \\
\label{sigmaeq}
0&=&-{1 \over 16\pi G}\e^{-2\phi}\left({\cal R} + 4\lambda^2 \right)
- - \left\{{a_1 \over 3}
\left({\cal R}+2\e^{2\sigma}\right)^2 + 4a_2\e^{2\sigma}{\cal R}\right\} \nn
&& + \e^{2\sigma}\left\{{4a_1 \over 3}\left({\cal R}+2\e^{2\sigma}\right)
+4a_2{\cal R}\right\} \nn
&& +{\e^{2\sigma} \over {\cal R}+2\e^{2\sigma}}\left\{{\beta_{a_1} \over 3}\left({\cal R}+2\e^{2\sigma}\right)^2 +4\beta_{a_2}\e^{2\sigma}{\cal R}\right\} \\
\label{rhoeq}
0&=&{1 \over 16\pi G}\e^{-2\phi}\left({\cal R}+2\e^{2\sigma} + 4\lambda^2
\right) + \left\{{a_1 \over 3}
\left({\cal R}+2\e^{2\sigma}\right)^2 + 4a_2\e^{2\sigma}{\cal R}\right\} \nn
&& - {\cal R}\left[{1 \over 16\pi G}\e^{-2\phi}
+{2a_1 \over 3}\left({\cal R}+2\e^{2\sigma}\right) +
4a_2{\cal R}\e^{2\sigma} \right.\nn
&& \left. +{1 \over 2\left({\cal R}+2\e^{2\sigma}\right) }
\left\{{\beta_{a_1} \over 3}\left({\cal R}+2\e^{2\sigma}\right)^2 +4\beta_{a_2}\e^{2\sigma}{\cal R}\right\}\right]
\eea
Consider that ${\cal R}$ is large but
constant. The constraint equations given by the variation over $g_{\pm\pm}$
are automatically satisfied if we assume that the solution is static and
spherically symmetric and ${\cal R}$, $\phi$ and $\sigma$ are constant
ones. Note that there is no consistent solution if we drop
the contribution from the quantum correction in Eqs.(\ref{phieq}),
(\ref{sigmaeq}) and (\ref{rhoeq}).( In the classical theory without
dilaton, there is, of course, a solution called the Nariai solution
where ${\cal R}$ and $\e^{2\sigma}$ are constant). The equation
(\ref{phieq}) given by the variation over the dilaton field $\phi$
forbids the existence of such a solution (with constant dilaton) at the
classical level.

There are some explicit solutions of (\ref{phieq}), (\ref{sigmaeq})
and (\ref{rhoeq}).
First we consider the case when the cosmological constant $\lambda^2$
is positive.
Then explicit solutions are given as
\bea
\label{solp}
\e^{2\sigma}&=&{3\lambda^2 \over 2}\left(-1+\sqrt{1
+{2\beta_{a_1} \over 3 \beta_{a_2}}
- -{\e^{-2\phi} \over 72\pi G \lambda^2 \beta_{a_2}}}\right)\ , \nn
{\cal R}&=&3\lambda^2 \left(-1-\sqrt{1
+{2\beta_{a_1} \over 3 \beta_{a_2}}
- -{\e^{-2\phi} \over 72\pi G \lambda^2 \beta_{a_2}}}\right)\ .
\eea
Here $\phi$ is given by the solution of the following equation:
\be
\label{solphi}
96\pi G \lambda^2 \left(\tilde a_1 + {\beta_{a_1} \over 2}\ln {G
|\lambda^2| \over \mu^2} \right) - 96\pi G \lambda^2 \beta_{a_1}\phi
=\e^{-2\phi}\ .
\ee
Eq.(\ref{solphi}) has two solutions if
\be
\label{condp1}
96\pi G \lambda^2 \left(\tilde a_1 + {\beta_{a_1} \over 2}\ln {G
|\lambda^2| \over \mu^2} \right)
+ 48\pi G \lambda^2 \beta_{a_1}\left\{\ln \left(
48\pi G \lambda^2 \beta_{a_1}\right) - 1 \right\}>0
\ee
and there is only one solution
\be
\label{solsp}
\phi=- {1 \over 2}\ln \left(48\pi G \lambda^2 \beta_{a_1}\right)
\ee
if
\be
\label{condp2}
96\pi G \lambda^2 \left(\tilde a_1 + {\beta_{a_1} \over 2}\ln {G
|\lambda^2| \over \mu^2} \right)
+ 48\pi G \lambda^2 \beta_{a_1}\left\{\ln \left(
48\pi G \lambda^2 \beta_{a_1}\right) - 1 \right\}=0\ .
\ee
The consistency condition that $\e^{2\sigma}$ in (\ref{solp}) is
positive is also given by
\be
\label{pconp}
{2\beta_{a_1} \over 3 \beta_{a_2}}
- -{\e^{-2\phi} \over 72\pi G \lambda^2 \beta_{a_2}}>0\ .
\ee
The conditions (\ref{condp1}) and (\ref{pconp}) are satisfied
independently. When the condition (\ref{pconp}) is satisfied, the
2d scalar curvature ${\cal R}$ in (\ref{solp}) is negative. The
4d scalar curvature is also negative: $R={\cal R} + 2\e^{2\sigma}
=-6\lambda^2<0$.

We now consider the case of $\lambda^2<0$. In this case, there are
two types of solutions. The first one corresponds to (\ref{solp}) and is
given by
\bea
\label{soln}
\e^{2\sigma}&=&{3\lambda^2 \over 2}\left(-1\pm\sqrt{1
+{2\beta_{a_1} \over 3 \beta_{a_2}}
- -{\e^{-2\phi} \over 72\pi G \lambda^2 \beta_{a_2}}}\right)\ , \nn
{\cal R}&=& 3\lambda^2 \left(-1\mp\sqrt{1
+{2\beta_{a_1} \over 3 \beta_{a_2}}
- -{\e^{-2\phi} \over 72\pi G \lambda^2 \beta_{a_2}}}\right)\ .
\eea
Here $\phi$ is given by the solution of (\ref{solphi}). The equation
(\ref{solphi}) has always only one solution when $\lambda^2<0$.
In order that $\e^{2\sigma}$ and ${\cal R}$ are real, the following
condition should be satisfied
\be
\label{pconn}
1+{2\beta_{a_1} \over 3 \beta_{a_2}}
- -{\e^{-2\phi} \over 72\pi G \lambda^2 \beta_{a_2}}>0\ .
\ee
If the condition (\ref{pconn}) is satisfied, $\e^{2\sigma}$ and ${\cal R}$
are always positive.
When $\lambda^2$ is negative, there is another type of solution
\be
\label{sols}
\e^{2\sigma}=-{3\lambda^2 \over 2}\ ,\ \ \
\e^{-2\phi}= 72\pi G \lambda^2 \left({2 \over 3}\beta_{a_1}
+ \beta_{a_2} \right)\ ,\ \ \ {\cal R}=-3\lambda^2 \ .
\ee
Note that $\e^{-2\phi}$ is consistently positive since
${2 \over 3}\beta_{a_1}+ \beta_{a_2}$ is negative.
In the solutions (\ref{soln}) and (\ref{sols}), 4d curvature is
also positive: $R={\cal R} + 2\e^{2\sigma}
=-6\lambda^2>0$.

When 2d scalar curvature ${\cal R}$ and $\e^{2\sigma}$ are positive
and constant ones as in  (\ref{soln}) and (\ref{sols}), the metric is
given by \be \label{rho0} ds^2= {2C\over {\cal R}} \cdot {1 \over
\cosh^2 \left(r\sqrt{C} \right)} \left(-dt^2 + dr^2 \right) +
\e^{2\sigma} d\Omega^2\ .  \ee Here $C$ is a positive constant of
integration. If we define new coordinate $\theta$ by \be \label{theta}
\sin\theta = \tanh \left(r\sqrt{C}\right) \ee the metric  becomes \be
\label{thetametric} ds^2={2C \over {\cal R}}\left(-\cos^2\theta dt^2 +
{1 \over C}d\theta^2\right) + \e^{2\sigma} d\Omega^2\ .
\ee
Since $\theta$ has period $2\pi$, the topology of three-dimensional
 space with
the metric (\ref{thetametric}) can be associated to
$S^1\times S^2$.
Note that according with Eq.(\ref{theta}) each
value of $r$ corresponds to two values of $\theta$.

When 2d scalar curvature ${\cal R}$ is negative
in (\ref{solp}), the metric is given by analytically continuing
$C \rightarrow -C$ in (\ref{rho0}):
\be
\label{rho0b}
ds^2= {2C\over {\cal R}}
\cdot {1 \over \cos^2 \left(r\sqrt{C} \right)}
\left(-dt^2 + dr^2 \right) + \e^{2\sigma} d\Omega^2\ .
\ee
The periodicy ${\pi \over \sqrt{C}}$ of the coordinate $r$ in the
metric (\ref{rho0b}) is not real but apparent one since there
are singularities in the metric when $r\sqrt{C}=\pm {\pi \over 2}$.
If we change the coordinate as
\be
\label{thetas}
\sinh s = \tan \left(r\sqrt{C}\right)
\ee
the metric  becomes
\be
\label{thetasmetric}
ds^2={2C \over {\cal R}}\left(-\cosh^2 s dt^2 +
{1 \over C}ds^2\right) + \e^{2\sigma} d\Omega^2\ .
\ee
According to (\ref{thetasmetric}) the topology of
the space is associated with
$R\times S^2$. The metric in the anti-de Sitter space can be expressed
by choosing the coordinates similar to those in (\ref{thetametric}):
\be
\label{AdSmetric}
ds^2={1 \over l^2}\left(-\cosh^2 s dt^2 +
ds^2\right) + {1 \over l^2}\sinh^2 s d\Omega^2\ .
\ee
Here $\Lambda=-3l^2$ is the cosmological constant.
Note that the essential difference between the above two metrics
(\ref{thetasmetric}) and (\ref{AdSmetric}) is that $\e^{2\sigma}$
is a constant or not.

The space-time with the metric (\ref{thetametric}) is known to be
the Nariai space. This solution is quite well-known in Einstein gravity with
cosmological constant. The Nariai space usually appears as the extremal
limit of Schwarzschild-de Sitter space, where the cosmological horizon of
de Sitter space and the horizon of the Schwarzschild black hole have the
same radius. Recently it has been shown the possibility that de
Sitter space (through creation of handles in the Nariai space) may be
disintegrated into several de Sitter universes \cite{B}.
Therefore the solutions
in (\ref{soln}) and (\ref{sols})
found here can be created by the quantum fluctuations
and may significally alter the cosmology of the early universe.
Therefore, we expect that handle creation takes place during inflation
not only due to primordial BHs production \cite{BH3} but also due to
quantum back-reaction of dilaton coupled matter as in above process.
The formed multiply BHs are responsible for disintegration of the universe
in the process of evaporation of these BHs. In other words, we presented
here new scenario to realize the inflationary universe (or, more exactly,
multiple inflationary universes) which appears in the process of
proliferation of de Sitter space.

In Eq.(\ref{XIV}), we assumed that $f(\phi)R$ is strong. 
In the inflationary universe, the magnitude of 
the scalar curvature $f(\phi)R$ is the 
square of Hubble constant $H$ : $f(\phi) R\sim H^2$ 
(usually the dilaton function $f(\phi)$ is absorbed into 
the redefinition of the metric tensor). 
The Hubble constant $H$ is given by $H^2\sim M^4 /m_{Pl}^2$.
Here $m_{Pl}$ is the Planck mass scale $10^{19}$ GeV and 
$M\sim\lambda$ is 
typically the GUT breaking mass scale $10^{14}$ GeV. 
Therefore $f(\phi) R$ would be about $10^{18}$ GeV$^2$, 
which is consistent in the solutions in (\ref{solp}),  
(\ref{soln}) or (\ref{sols}) (and (\ref{solphi})). 
Therefore $f(\phi) R$ could be sufficiently strong. If we choose 
$\mu\sim M$, we find $t={1 \over 2}\ln {|f(\phi) R| \over \mu^2}\sim 10$ 
the effective coupling constants would be $a_1$, $a_2\sim 10^{-2}\times N$.

It is very interesting to
note that in the same way one can generate other
well-known classical solutions as the result
of quantum back-reaction of dilaton coupled matter. For example,
for non-zero electromagnetic field one can study the possibility
for quantum generation of charged S(anti)dS BHs (for classical
description of charged Nariai BHs, see \cite{BO}).
It might be also interesting to investigate the possibility of the
disintegration especially in the case of the solution with negative
curvature.

Let us consider the
 solution (\ref{sols}) with the negative curvature.
 When the solution is perturbed, $\e^{2\sigma}$ would decrease
and vanish in some regions in the space described
by the metric (\ref{thetasmetric}).
These regions horizons would correspond to the black hole horizons
in Schwarzschild-de Sitter black hole
although the solution (\ref{sols}) would not correspond to
 any extremal limit of black hole.
On the other hand, in other regions with the cosmological horizons,
$\e^{2\sigma}$ would increase and go to infinity by the perturbation.
Therefore as in the disintegration from the Nariai space to de Sitter space \cite{B},
the space is also disintegrated into several universes.
The generated universes are large enough since the
radius coordinate $\e^{2\sigma}$ could
become very big in the regions corresponding to the cosmological horizon.
Hence the universe could be described by the classical theory and it should 
be anti-de Sitter
space with metric (\ref{AdSmetric}). Therefore the above
process describes the proliferation of anti-de Sitter space.

\

\noindent
{\bf Acknoweledgments} SDO is grateful to S. Hawking for useful
 discussion. This
work has been partially supported by RFBR,project n96-02-16017 and
GRACENAS (No.6-18-1997) (Russia).  AAB thanks CNPq (Brasil) for
 financial support.


\begin{thebibliography}{99}
\bibitem{B} R. Bousso, hep-th/9805081.
\bibitem{GP} P. Ginsparg and M.J. Perry, {\it Nucl.Phys.}
 {\bf B222} (1983) 245.
\bibitem{NO} S. Nojiri and S.D. Odintsov, 
{\it Phys.Lett.} {\bf B426} (1998) 29.
\bibitem{IO} S. Ichinose and S.D. Odintsov, hep-th/9802043.
\bibitem{BOS} I.L. Buchbinbder, S.D. Odintsov and I.L.
Shapiro, {\sl Effective Action in Quantum Gravity},
IOP Publishing, Bristol and Philadelphia, 1992.
\bibitem{CW} S. Coleman and E. Weinberg, {\it Phys.Rev.}
{\bf D7} (1973) 1888;
M.B. Einhorn and D.R.T. Jones, {\it Nucl.Phys.}
{\bf B211} (1983) 29; {\bf B230} (1984) 261;
G.B. West, {\it Phys.Rev.} {\bf D27} (1983) 1402;
K. Yamaguchi, {\it Nucl.Phys.} {\bf B216} (1983) 508;
C. Ford, D.R.T. Jones, P.W. Stephenson and M.B. Einhorn,
{\it Nucl.Phys.} {\bf B395} (1993) 17.
\bibitem{BH2} R. Bousso and S.W. Hawking, {\it Phys.Rev.}
{\bf D57} (1998) 2436.
\bibitem{NO2} S. Nojiri and S.D. Odintsov, hep-th/9802160,
hep-th/9804033.
\bibitem{BH3} R. Bousso and S.W. Hawking, {\it Phys.Rev.}
 {\bf D54} (1996) 6312; {\bf D52} (1995) 5659.
\bibitem{BO} R.Bousso, {\it Phys.Rev.} {\bf D55} (1997) 3614.
\end{thebibliography}
\end{document}